\documentclass[showpacs,preprintnumbers,amsmath,amssymb,prb,twocolumn]{revtex4}

\usepackage{graphicx}% Include figure files
\usepackage{dcolumn}% Align table columns on decimal point
\usepackage{bm}% bold math
\def\comment#1{}

\def\slashchar#1{\setbox0=\hbox{$#1$}           % set a box for #1 
   \dimen0=\wd0                                 % and get its size
   \setbox1=\hbox{/} \dimen1=\wd1               % get size of /
   \ifdim\dimen0>\dimen1                        % #1 is bigger
      \rlap{\hbox to \dimen0{\hfil/\hfil}}      % so center / in box
      #1                                        % and print #1
   \else                                        % / is bigger
      \rlap{\hbox to \dimen1{\hfil$#1$\hfil}}   % so center #1
      /                                         % and print /
   \fi}                                         %

\def\sigmab{{\mbox{\boldmath $\sigma$}}}

\begin{document}

\title{Deconfined quantum criticality driven by Dirac fermions in 
SU(2) antiferromagnets}

\author{Flavio S. Nogueira}
\email{nogueira@physik.fu-berlin.de}
\affiliation{Institut f{\"u}r Theoretische Physik,
Freie Universit{\"a}t Berlin, Arnimallee 14, D-14195 Berlin, Germany}

\date{Received \today}

\begin{abstract}
Quantum electrodynamics in 2+1 dimensions is an effective gauge 
theory for the so called algebraic quantum liquids. A new type 
of such a liquid, the algebraic charge liquid, has been proposed 
recently in the context of deconfined quantum critical 
points [R. K. Kaul {\it et al.}, Nature Physics {\bf 4}, 28 (2008)]. 
In this context, we show by using the renormalization group in $d=4-\epsilon$ spacetime 
dimensions, that a deconfined quantum critical point 
occurs in a SU(2) system provided the number of Dirac fermion species 
$N_f\geq 4$. The calculations are done in a representation where the 
Dirac fermions are given by four-component spinors. The critical exponents are calculated  
for several values of $N_f$. In particular, for $N_f=4$ and 
$\epsilon=1$ ($d=2+1$) the anomalous dimension of the N\'eel field 
is given by $\eta_N=1/3$, with a correlation length exponent 
$\nu=1/2$.  These values change considerably for $N_f>4$. For instance, 
for $N_f=6$ we find $\eta_N\approx 0.75191$ and $\nu\approx 0.66009$. We also investigate 
the effect of chiral symmetry breaking and analyze the 
scaling behavior of the chiral holon susceptibility, 
$G_\chi(x)\equiv\langle\bar \psi(x)\psi(x)\bar \psi(0)\psi(0)\rangle$.
\end{abstract}

\maketitle

\section{Introduction}

In very simple terms we can define 
spinons as elementary excitations carrying spin but no charge. They emerge 
in two-dimensional Mott insulators as a consequence of electron fractionalization.
\cite{Sachdev-Review}  
Similarly to quarks in QCD, spinons do not emerge so easily out of a confined 
state. As a consequence, {\it explicit} observation of spinons is very 
difficult. Quarks, though being confined, are essential to 
explain the properties and structure of matter. In a similar way, 
spinons, in spite of their confinement, seem to be essential in explaining 
all the remarkable properties of Mott insulators, including the doped ones, opening 
the way for the understanding of high-$T_c$ superconductors.\cite{Anderson} 

In several systems spinons may deconfine at a quantum critical point.\cite{Sachdev-Review} 
One prominent example is the critical point separating different types of order 
in quantum antiferromagnets, like for example in the 
phase transition between a N\'eel and a valence-bond solid (VBS) state,\cite{RS} which is 
a paramagnetic state breaking lattice symmetries.  
Spinon deconfinement also occurs in some 
paramagnetic Mott insulating states where no symmetries are broken, like for example the spin liquid   
state in pyrochlore antiferromagnets,\cite{Moessner,Nussinov} and in some theories for the 
underdoped state of high-$T_c$ superconductors.\cite{Wen-RMP}
A remarkable property of deconfined spinons is that they are interacting at low energies 
(i.e., they are aymptotic interacting in the infrared), in contrast with deconfined quarks, which are free at high energies  
(asymptotic freedom).\cite{AF} Therefore, the field theory of a deconfined 
quantum critical point is a highly nontrivial conformal field theory (CFT).
\cite{Senthil-2004} This deconfined quantum criticality (DQC) arises due to a 
destructive interference between the Berry phases of a quantum antiferromagnet and 
the instantons. This mechanism makes instanton events irrelevant at large distances, 
which in turn allows the spinons to deconfine. Furthermore, since  
deconfined spinons are strongly interacting, the critical exponents are nontrivial. 
Interestingly, the critical exponents following from such a theory have unusual 
values with respect to those arising from the Landau-Ginzburg-Wilson (LGW) paradigm 
of phase transitions.\cite{Senthil-2004} The difference can already be seen in mean-field 
theory. It is easy to see that mean-field theory based on the LGW paradigm gives 
an anomalous dimension $\eta_N=0$ for the N\'eel field ${\bf S}_i=(-1)^iS{\bf n}_i$, 
where $i$ is the lattice site, 
$S$ is the spin, and ${\bf n}_i^2=1$. However, this is not the case 
for a mean-field theory 
in the DQC scenario. There the direction field is written in terms of the 
spinon fields using a CP$^1$ 
representation, i.e., $n_{i,a}={\bf z}_i^\dagger\sigma_a{\bf z}_i$, $a=1,2,3$, 
where ${\bf z}_i=(z_{i1},z_{i2})$, and $\sigma_a$ are the Pauli matrices, and  
the constraint ${\bf n}_i^2=1$ implies ${\bf z}_i^\dagger{\bf z}_i=1$. 
In mean-field theory the dimension of fields is simply given by dimensional analysis of the 
action functional. In such a case this gives simply 
${\rm dim}[z_{i\alpha}]=(d-2)/2$, where $d$ is the dimension of the spacetime. 
Therefore, mean-field theory leads to 
${\rm dim}[n_{i,a}]=(d-2+\eta_N)/2=2{\rm dim}[z_\alpha]$, which 
implies $\eta_N=d-2$, or $\eta_N=1$ in $2+1$ dimensions. Thus, we see that 
the field theory of deconfined spinons leads to a large anomalous dimension of 
the N\'eel field already at the mean-field level. Quantum fluctuations typically 
reduce this value, but it still remains 
considerably larger than the prediction of the LGW scenario,
\cite{Motrunich,Sandvik} 
whose fluctuation 
corrections lead to an anomalous dimension $\eta_N\approx 0.03$.\cite{KSF}

A deconfined quantum critical point governs a second-order phase transition between 
different quantum states of matter at zero temperature, the paradigmatic 
situation in this case being the transition from a N\'eel state to 
a VBS. Thus, it is important to check whether  
the proposed models for DQC actually undergo a second-order phase transition with a 
large anomalous dimension. There has been in recent years a lot of activity in this direction. 
Indeed, some of the proposed models have been checked extensively in large scale Monte Carlo (MC) 
simulations.
\cite{Motrunich,Sandvik,Kuklov_2006,Kragset,Harada,Melko,Jiang} Particularly important are the 
MC simulations for the well studied 
$S=1/2$ quantum antiferromagnet with easy-plane anisotropy. Due 
to its global $U(1)$ symetry and the additional {\it local} $U(1)$ symmetry, the model 
is self-dual.\cite{Motrunich} This model was predicted to exhibit a deconfined quantum critical 
point.\cite{Motrunich,Senthil-2004} However, recent MC simulations \cite{Kuklov_2006,Kragset} 
have clearly shown that in this case actually a first-order phase transition takes place. This 
result is confirmed by a recent renormalization group (RG) analysis.\cite{Nogueira-Kragset-Sudbo} 
Remarkably, despite the absence of quantum criticality, the destructive interference mechanism 
between the Berry phases and instantons was shown to work,\cite{Kragset} so that we can 
still talk about spinon deconfinement. The situation here is reminiscent of the 
$Z_2$-Higgs theory in three spacetime dimensions, though in this case we have a discrete gauge 
symmetry rather than a continuous one. This model is also self-dual,\cite{Balian} but  
is known to undergo a first-order phase transition along the self-dual line.\cite{Stack} 
However, the $Z_2$-Higgs theory has the peculiarity of having a deconfined phase 
bounded by two lines of second-order phase transitions that become first-order when they 
are very near each other and to the self-dual line. 
It is not known at present whether a similar behavior  
for the easy-plane frustrated quantum antiferromagnet is possible. 

For the $SU(2)$ case, a MC simulation on a Heisenberg model with hedgehog 
suppression \cite{Motrunich} seems to support the DQC scenario. Another interesting model is the 
$S=1$ Heisenberg antiferromagnet with a biquadratic interaction between the spins,\cite{Grover}  
where numerical evidence for DQC has also been found.\cite{Harada} 
Strong evidence for DQC has been reported in Ref. \onlinecite{Sandvik}, where a model featuring a
four-spin interaction was simulated. There the anomalous dimension of the N\'eel field 
was found to be $\eta_N\approx 0.26$. Further MC studies \cite{Melko} confirm 
the analysis of Ref. \onlinecite{Sandvik}, in which a transition from a N\'eel state 
to a VBS occurs. There the obtained value of the anomalous dimension is 
$\eta_N\approx 0.35$. However, there is a recent paper 
\cite{Jiang} on the same model where MC simulations are reported to lead to a first-order 
phase transition. 

Due to the inherently non-perturbative character of deconfined quantum critical points, purely analytical and 
well controlled studies are not easy to perform. 
For instance, the perturbative RG applied 
to the $SU(N)$ case can only access a quantum critical point if $N>182.9$.\cite{Nogueira-Kragset-Sudbo} 
The smallest value of $N$ producing a critical point, $N=183$, leads to an anomalous 
dimension $\eta_N=609/671\approx 0.9076$. In principle the result of this RG analysis would mean that the phase transition 
for the $SU(2)$ case is a first-order one. However, it was argued in Ref. \onlinecite{Nogueira-Kragset-Sudbo} that, similarly to 
the Ginzburg-Landau (GL) superconductor with $N$ complex order parameter fields,\cite{Halperin-Lubensky-Ma} the actual critical 
behavior at low $N$ and in the strong coupling limit may be compatible with a second-order phase transition. For 
a GL model with a single complex order field, this expectation is confirmed both by a duality and MC analysis of the model.
\cite{Dasgupta,Kleinert-tric,Kiometzis,Herbut,Olsson,Hove,Neuhaus} There are RG studies where low-$N$ critical points 
were found,\cite{Berg,Folk,Herbut-Tesanovic,Kleinert-Nogueira-1} but they all have some kind of problem, being either 
not well controlled or involving ad hoc assumptions. The two-loop perturbation in terms of the $\epsilon$-expansion  
should be in principle a well controlled approach.\cite{Kolnberger} The resummation of the two-loop result \cite{Folk} 
led to a critical point for $N=1$. However, the unresummed result, though correct, contains unacceptable pathologies, 
like for example a critical exponent $\nu$ for $182.9<N\leq 200$ larger than the $N\to\infty$ result at fixed 
dimension, and the absence of a fixed point for the gauge coupling if $N<18\epsilon$. 
In view of these pathologies, the resummed two-loop result cannot be completely trusted. The ideal approach would be to 
obtain a resummed three-loop result. Unfortunately, to this order the only available RG function is the $\beta$ function for 
the gauge coupling.\cite{3-loop} Anyway, the important argument from Ref. \onlinecite{Nogueira-Kragset-Sudbo} to be retained 
here is the following. The existence of a critical value of $N$ in the 
weak coupling analysis in $d=4-\epsilon$ spacetime dimensions, above which a critical point exists, is thought as an 
indication that in the strong-coupling limit a quantum critical point at lower values of $N$ may exist. This is of course 
a conjecture that must be tested further. Note, however, that the easy-plane case, also studied in 
Ref. \onlinecite{Nogueira-Kragset-Sudbo}, when generalized to a theory with a global $O(N)\times O(N)$ symmetry, does 
not have any fixed points at nonzero gauge coupling for all values of $N$, i.e., this theory does not have a critical 
value of $N$.       

The behavior of the $\epsilon$-expansion for the GL model with a single complex order field improves considerably if 
the theory is coupled via the gauge field to $N_f$ Dirac fermions species.\cite{Kleinert-Nogueira-2} Interestingly, 
it is not necessary to have a large value of $N_f$ to obtain an infrared stable fixed point. In the context of the 
present paper, it should be noted that the $SU(2)$ model for deconfined spinons is exactly the same as a GL model 
with two complex order fields. 

In this paper we will consider the Lagrangian:

\begin{equation}
\label{L}
{\cal L}={\cal L}_b+{\cal L}_f,
\end{equation}
where
 
\begin{eqnarray}
\label{L-boson}
{\cal L}_b&=&\frac{1}{2e_0^2}(\epsilon_{\mu\nu\lambda}\partial_\nu A_\lambda)^2+
\sum_{\alpha=1}^{N_b}|(\partial_\mu-iA_\mu)z_\alpha|^2\nonumber\\
&+&r_0\sum_{\alpha=1}^{N_b}|z_\alpha|^2+\frac{u_0}{2}
\left(\sum_{\alpha=1}^{N_b}|z_\alpha|^2\right)^2,
\end{eqnarray}
and

\begin{equation}
\label{L-fermion}
{\cal L}_f=\sum_{\alpha=1}^{N_f}\bar \psi_a(\slashchar{\partial}+i\slashchar{A})\psi_a.
\end{equation}
The Lagrangian ${\cal L}_b$ is precisely the model for deconfined spinons proposed in Ref. 
\onlinecite{Senthil-2004}. In this context, we are primarily interested in the case with $N_b=2$. 
The Lagrangian ${\cal L}_b$ is supposed to govern the universality class of the phase transition between 
a N\'eel state and a VBS. 

The Lagrangian ${\cal L}_f$ contains $N_f$ species of four-component Dirac fermions in $2+1$ 
Euclidean dimensions. 
\cite{Pisarski}  The representation of $\gamma$ matrices in this case is given by

\begin{equation}
\gamma_0=\left(
\begin{array}{cc}
\sigma_3 & 0\\
\noalign{\medskip}
0 & -\sigma_3
\end{array}
\right),~~~~~~~~~
\gamma_1=\left(
\begin{array}{cc}
\sigma_2 & 0\\
\noalign{\medskip}
0 & -\sigma_2
\end{array}
\right),
\nonumber
\end{equation}
\begin{equation}
\gamma_2=\left(
\begin{array}{cc}
\sigma_1 & 0\\
\noalign{\medskip}
0 & -\sigma_1
\end{array}
\right),
\end{equation}
where $\sigma_1$, $\sigma_2$, and $\sigma_3$ are the Pauli matrices.  

The Lagrangian ${\cal L}_f$ is typical of a so called algebraic quantum liquid. There are three such quantum liquids, 
the algebraic spin liquid (ASL),\cite{Rantner} the algebraic Fermi liquid (AFL),\cite{FT} and the recently introduced 
algebraic charge liquid (ACL).\cite{Kaul} Let us briefly mention the differences between these algebraic quantum liquids. 

The ASL emerges out of the so called staggered flux phase \cite{Affleck} in the large $N_f$ limit 
of a $SU(N_f)$ 
Heisenberg antiferromagnet. In this case the spin operators are written in terms of fermion bilinears such that the 
Heisenberg model reads $H=-(J/N_f)\sum_{\langle i,j\rangle}f_{i\alpha}^\dagger f_{j\alpha}f_{j\beta}^\dagger f_{i\beta}$, 
with the constraint $f_{i\alpha}^\dagger f_{i\alpha}=N_f/2$. A lattice gauge field arises from the phase fluctuations of 
the Hubbard-Stratonovich link field $\chi_{ij}=\langle f_{i\alpha}^\dagger f_{j\alpha}\rangle$. This gives rise to a 
compact $U(1)$ gauge theory. Linearizing around the nodes $\pm(\pi/2,\pi/2)$ of the quasi-paricle spectrum 
$E_{\bf k}=2|\chi_0|\sqrt{\cos^2k_x+\cos^2k_y}$ and taking the continuum limit leads to the Lagrangian (\ref{L-fermion}). 
For large enough $N_f$ the fermionic spinons deconfine \cite{Hermele,Nogueira-Kleinert-1,Nogueira-Kleinert-2} 
and we obtain a Mott insulating state having no broken symmetries, 
i.e., a spin liquid. For low enough $N_f$ the spin liquid is no longer stable \cite{Nogueira-Kleinert-2} and 
the chiral symmetry is probably broken due to spinon confinement. However, in this theory chiral symmetry breaking 
(CSB) takes place even in the non-compact case.\cite{Pisarski,Appelq} This CSB is associated to the development 
of an insulating antiferromagnetic state.\cite{Kim}

The AFL, though having essentially the same Lagrangian ${\cal L}_f$, is physically very different 
from the ASL. Here the gauge field is not compact and has a completely different origin.\cite{FT,Herbut-AFL} 
While the 
ASL is associated to a Mott insulator, the AFL originates from a $d$-wave superconductor. The Dirac fermions 
are obtained by identifying the low-energy quasi-particles on the nodes of the $d$-wave gap. The gauge field 
follows from the coupling of these quasi-particles to vortices. The way this is done is very subtle and 
involves the manipulation of singular gauge transformations.\cite{FT} The fermions in the AFL, just like the ones 
of the ASL, carry no charge. Interestingly, the number $N_f$ in the AFL is related to the number of 
layers of the system. For example, a single-layer system has $N_f=2$, while a bilayer system has $N_f=4$. Note that 
in this case there is no need for other gauge fields in the corresponding layers, since the gauge field arises from 
a vortex-antivortex excitation. Thus, at large distances a vortex in one layer is always connected with a vortex 
in the second layer, the same happening for the antivortices. The end result is that Dirac fermions in different layers 
couple to the same gauge field.\cite{Note-1} Of course, similarly to the ASL,  CSB also occurs here 
for small enough $N_f$,\cite{Herbut-AFL,Zlatko-CSB} and is once more associated to antiferromagnetism.        
 
In the ACL a fractional fermionic particle with charge $e$ and no spin, a {\it holon},  
couples to bosonic spinons via a gauge field.\cite{Kaul} Therefore, it is necessarily related 
to the concept of DQC and in this case  
the whole Lagrangian (\ref{L}) has to be considered. 
Note that in contrast with the ASL and AFL, the Dirac fermions in the ACL carry charge. This 
new state of matter bears some resemblances with earlier ideas on spin-charge separation in 
the cuprates.\cite{Wen-RMP} An important difference is that in earlier theories superconductivity 
is obtained by doping a spin liquid, while in the ACL superconductivity arises by doping an antiferromagnetic 
state near the deconfined quantum critical point.\cite{Kaul-PRB} Thus, we can either dope a N\'eel or a 
VBS state. Like in the two other algebraic quantum liquids, here the chiral symmetry can also be broken. However, 
since the Dirac fermions now carry charge, we have a situation where charge density waves develop as a result of 
CSB. It is important to note that the coupling to bosons leads to a reduction of the value of $N_f$ below which 
the chiral symmetry breaks.\cite{Kim} 

From the above discussion we see that among the three algebraic quantum liquids, only the ACL cannot exist 
without the coupling to bosons. The results of this paper will concern mainly this case. Quantum criticality 
in $U(1)$ gauge theories with both bosonic and fermionic matter has been also considered in a 
recent paper.\cite{Kaul-Sachdev} There the authors considered a CP$^{N_b-1}$ model coupled to 
$N_f$ Dirac fermion species and analyzed the model for large $N_b$ and $N_f$, while 
keeping $N_f/N_b$ arbitrary. Here we have 
followed Ref. \onlinecite{Senthil-2004} and softened the CP$^{N_b-1}$ constraint. This leads to a 
theory that is tractable in a RG framework in $d=4-\epsilon$ spacetime dimensions. The main advantage  
of this approach is that we will be able to work with a fixed $N_b=2$ and $N_f$ does not need to be large to 
obtain a quantum critical point. Part of our results follow directly from Ref. 
\onlinecite{Kleinert-Nogueira-2}, the main difference being that here we are interested in 
$N_b=2$ rather than $N_b=1$.\cite{Note-2} However, in this paper we will improve substantially upon the 
previous analysis and present many new results which are relevant for the quantum critical behavior of 
the ACL. One of the main new results of this paper will be the 
calculation of $\eta_N$ for $N_b=2$ and $N_f\geq 4$ in terms of the crossover exponent $\varphi$. As shown in 
Ref. \onlinecite{Nogueira-Kragset-Sudbo}, the critical exponent $\eta_N$ is related to the crossover 
exponent by the formula

\begin{equation}
\label{eta-N}
\eta_N=d+2(1-\varphi/\nu),
\end{equation}
where $\nu$ is the correlation length exponent. The crossover exponent is related to a mass anisotropy of 
the Lagrangian. Although the Lagrangian ${\cal L}_b$ does not have any mass anisotropy, the correlation 
function $\langle{\bf z}^*(x)\cdot{\bf z}(0)~{\bf z}(x)\cdot{\bf z}^*(0)\rangle$ is calculated through 
the insertion of the operator $z_\alpha^*z_\beta$. For $\alpha\neq\beta$ this insertion allows us to 
calculate $\varphi$.\cite{Amit} 

The plan of the paper is as follows. In order to 
explain why a second-order phase transition is easier 
to obtain when Dirac fermions are included, in Section II we will calculate the effective potential associated 
to the Lagrangian (\ref{L}). There we compare the cases with and without fermions. In the absence of 
fermions the one-loop effective potential typically describes a first-order phase transition. When 
fermions are included the situation changes and we can show that for small $N_f$ a first-order transition 
takes place, while for larger values of $N_f$ the effective potential features a second-order phase 
transition. In Section III we proceed with the RG analysis and the calculation of the critical exponents. 
In Section IV we briefly discuss the effect of CSB and show that it does not affect the quantum critical 
regime analyzed in Section III. We also compute the anomalous dimension of the chiral holon susceptibility. 
Section V concludes the paper. An Appendix sketches the details of the calculations of the 
anomalous dimension $\eta_N$.   

\section{Effective potentials and the order of the phase transition}

\subsection{A simple example}

In order to illustrate the subtlety involving the determination of the order of the phase transition in models 
for deconfined spinons, we will consider the following simple Lagrangian for an interacting scalar theory 
in $d=2+1$ Euclidean dimensions:

\begin{equation}
{\cal L}=\frac{1}{2}(\partial_\mu\phi)^2+\frac{m_0^2}{2}\phi^2+\frac{u_0}{4!}\phi^4.
\end{equation}

At the mean-field level the above model obviously exhibits a second-order phase transition with a 
critical point at $m_0^2=0$.  
Let us calculate the effective potential for the above model at one-loop order. This is more easily done 
by writing $\phi=\bar \phi+\delta\phi$, where 
$\bar \phi$ is a constant background field,  and integrating out the quadratic fluctuations in $\delta\phi$ while 
disregarding the higher order ones. This calculation is very simple, since it just involves a Gaussian 
integration in $\delta\phi$. We obtain, 

\begin{eqnarray}
&U_{\rm eff}(\bar \phi)=\frac{m_0^2}{2}\bar \phi^2+\frac{u_0}{4!}\bar \phi^4
\nonumber\\
&+\frac{1}{2V}\left[{\rm Tr}\ln\left(-\partial^2+m_0^2+\frac{u_0}{2}\bar \phi^2\right)
-{\rm Tr}\ln(-\partial^2)\right],
\end{eqnarray}
where $V$ is the (infinite) volume. Explicit evaluation of the tracelog term yields

\begin{equation}
U_{\rm eff}(\bar \phi)=\frac{1}{2}\left(m_0^2+\frac{\Lambda u_0}{2\pi^2}\right)\bar \phi^2
+\frac{u_0}{4!}\bar \phi^4-\frac{1}{12\pi}\left(m_0^2+\frac{u_0}{2}\bar \phi^2\right)^{3/2},
\end{equation}
where $\Lambda$ is a ultraviolet cutoff and we have neglected terms which are independent from 
$\bar \phi$. The above effective potential can be written in terms of renormalized quantities $m^2$ and 
$u$ defined by the normalization conditions $U''_{\rm eff}(0)=m^2$ and 
$U''''_{\rm eff}(0)=u$, where the primes denote derivatives with respect to $\bar \phi$. This leads 
to 

\begin{equation}
m^2=m_0^2+\frac{\Lambda u_0}{2\pi^2},
\end{equation}
and
\begin{equation}
\label{u}
u=u_0-\frac{3u_0^2}{16\pi m},
\end{equation}
where in Eq. (\ref{u}) we have replaced in the second term $m_0^2$ by $m^2$, since the error involved 
in this replacement contributes to an order higher than the one being calculated here. Thus, the 
effective potential becomes

\begin{equation}
U_{\rm eff}(\bar \phi)=\frac{m^2}{2}\bar \phi^2
+\frac{u_0}{4!}\bar \phi^4-\frac{1}{12\pi}\left(m^2+\frac{u_0}{2}\bar \phi^2\right)^{3/2}.
\end{equation}  
The above potential is typical of a system exhibiting a second-order phase transition. The order parameter 
is obtained by extremizing the effective potential. It is given by

\begin{equation}
\bar \phi_\pm=\pm\frac{\sqrt{3}}{8\pi u_0}\left(3u_0^3-128\pi^2m^2u_0-\sqrt{9u_0^6-512\pi^2m^2u_0^4}\right)^{1/2},
\end{equation}
and we see that the critical point is given by $m^2=0$. Looking at Eq. (\ref{u}) we may think that 
$u$ will get large and negative as the critical point is approached, such that the effective potential would 
become unstable against a $\phi^6$ interaction. However, this is not the case, since up to one-loop accuracy we 
can write the following RG equation for the dimensionless coupling $\hat u=u/(16\pi m)$, 

\begin{equation}
m\frac{\partial\hat u}{\partial m}=-\hat u+3\hat u^2.
\end{equation}
We see that as $m\to 0$ the above $\beta$ function vanishes at the infrared stable fixed point 
$\hat u_*=1/3$. Thus, $u$ actually does not diverge as the critical point is approached.  

\subsection{Effective potential for the $SU(2)$ antiferromagnet}

Let us consider now the one-loop effective potential for the Lagrangian (\ref{L-boson}). Like in the example 
of a single scalar field theory, all we have to do is to integrate out the Gaussian fluctuations. Note that 
the Lagrangian is already quadratic in the gauge field, so that the latter can be immediately integrated 
out exactly. We will consider a nonzero background for the fields $z_1$ and $z_2$, leaving a vanishing 
background for the remaining $N_b-2$ fields. Thus, we will have $z_1=\bar z_1+\delta z_1$, 
$z_2=\bar z_2+\delta z_2$, and $z_\alpha=\delta z_\alpha$ for $\alpha\geq 3$. Thus, we have for 
$N_b=2$ the background spin orientation field $\bar {\bf n}=\bar z_\alpha\sigmab_{\alpha\beta}\bar z_\beta$. 
The result is

\begin{widetext}
\begin{eqnarray}
\label{Ueff-AF}
&U_{\rm eff}^{\rm AF}(\bar z_1,\bar z_2)=m^2(|\bar z_1|^2+|\bar z_2|^2)+\frac{u_0}{2}(|\bar z_1|^2+|\bar z_2|^2)^2
-\frac{\sqrt{2}e_0^3}{3\pi}\left(|\bar z_1|^2+|\bar z_2|^2\right)^{3/2}\nonumber\\
&-\frac{1}{12\pi}\left\{\left[m^2+u_0\left(3|\bar z_1|^2+|\bar z_2|^2\right)\right]^{3/2}
+\left[m^2+u_0\left(|\bar z_1|^2+3|\bar z_2|^2\right)\right]^{3/2}
+2(N_b-1)\left[m^2+u_0\left(|\bar z_1|^2+|\bar z_2|^2\right)\right]^{3/2}\right\},
\nonumber\\
\end{eqnarray}
%\end{widetext}
where $m^2$ is given by

\begin{equation}
\label{m2}
m^2=r_0+\left(\frac{N_b+1}{2}u_0+\frac{4e_0^2}{3}\right)\frac{\Lambda}{\pi^2}.
\end{equation} 
The third term in Eq. (\ref{Ueff-AF}) arises from the integration over the gauge field and is reminiscent 
from the fluctuation-corrected mean-field theory in the GL superconductor.
\cite{Halperin-Lubensky-Ma,Nogueira-Kleinert-Lecture} Due to this term the symmetry will be broken for 
a $m^2>0$, leading in this way to a weak first-order phase transition.
\cite{Halperin-Lubensky-Ma} Note that for $e_0=0$ we have a second-order phase transition in the universality class 
of a $O(2N_b)$ symmetric classical magnetic system in three dimensions. 

\subsection{Effective potential for the algebraic charge liquid}                    

Finally, let us consider the the one-loop effective potential for the the Lagrangian (\ref{L}). 
First we note that the one-loop contribution following from integrating out the Dirac fermions is 
given by

\begin{equation}
{\cal L}_f^{\rm eff}=\frac{N_fe_0^2}{32}F_{\mu\nu}\frac{1}{\sqrt{-\partial^2}}F_{\mu\nu},
\end{equation}
where $F_{\mu\nu}=\partial_\mu A_\nu-\partial_\nu A_\mu$ (note that the above expression is 
multiplied by a factor two in the case of two-component Dirac fermions). 
The above expression follows simply by calculating the 
one-loop vacuum polarization in $2+1$ dimensions.  The background fields are chosen in the 
same way as in the previous Subsection. After integrating out the quadratic gauge fluctuations
for a fixed spinon background, we obtain
the correction

\begin{equation}
\delta{\cal L}_f^{\rm eff}=\frac{1}{V}\left\{{\rm Tr}\ln\left[-\partial^2+\frac{N_fe_0^2}{8}\sqrt{-\partial^2}+2e_0^2
(|\bar z_1|^2+|\bar z_2|^2)\right]-{\rm Tr}\ln(-\partial^2)\right\}.
\end{equation}
After evaluating the above tracelog and integrating out the spinon Gaussian fluctuations, we obtain 

%\begin{widetext}
\begin{eqnarray}
\label{Ueff-ACL}
&&U_{\rm eff}^{\rm ACL}(\bar z_1,\bar z_2)=m^2
(|\bar z_1|^2+|\bar z_2|^2)+\frac{u_0}{2}(|\bar z_1|^2+|\bar z_2|^2)^2
%+\left(\frac{u_0}{2}-\frac{128\Lambda}{3\pi^2N_f^2}\right)(|\bar z_1|^2+|\bar z_2|^2)^2
%\nonumber\\
+\frac{N_fe_0^4}{16\pi^2}\left(|\bar z_1|^2+|\bar z_2|^2-\frac{N_f^2e_0^2}{384}\right)
\ln\left[\frac{2e_0^2(|\bar z_1|^2+|\bar z_2|^2)}{\Lambda^2}\right]
\nonumber\\
&&+\frac{(N_fe_0^2/8)^3-(5N_fe_0^4/4)(|\bar z_1|^2+|\bar z_2|^2)+(128e_0^2/N_f)(|\bar z_1|^2+|\bar z_2|^2)^2}{
12\pi^2\sqrt{1-\frac{512}{N_f^2e_0^2}(|\bar z_1|^2+|\bar z_2|^2)}}\ln\left[\frac{1-\sqrt{1-\frac{512}{N_f^2e_0^2}(|\bar z_1|^2+|\bar z_2|^2)}}{1+\sqrt{1-\frac{512}{N_f^2e_0^2}(|\bar z_1|^2+|\bar z_2|^2)}}\right]
\nonumber\\
%&-&\frac{2048}{3\pi^2N_f^3}(|\bar z_1|^2+|\bar z_2|^2)^3\ln\left[\frac{16(|\bar z_1|^2+|\bar z_2|^2)}{N_f\Lambda}
%\right]\nonumber\\
&&-\frac{1}{12\pi}\left\{\left[m^2+u_0\left(3|\bar z_1|^2+|\bar z_2|^2\right)\right]^{3/2}
+\left[m^2+u_0\left(|\bar z_1|^2+3|\bar z_2|^2\right)\right]^{3/2}
+2(N_b-1)\left[m^2+u_0\left(|\bar z_1|^2+|\bar z_2|^2\right)\right]^{3/2}\right\},\nonumber\\
\end{eqnarray}
%\end{widetext}
where $m^2$ is still given by Eq. (\ref{m2}). 
Note that the annoying third term of Eq. (\ref{Ueff-AF}) disappeared once Dirac fermions were included. For  
$N_f\to 0$ the above equation reduces to Eq. (\ref{Ueff-AF}), as it should.  Thus, we can expect that for $N_f$ sufficiently 
small a first-order phase transition occurs. On the other hand, for $N_f$ above some critical value a second-order phase transition 
should take place. The leading small $N_f$ correction shifts the mass term and add a logarithmic correction, resulting in 

\begin{equation}
U_{\rm eff}^{\rm ACL}(\bar z_1,\bar z_2)=U_{\rm eff}^{\rm AF}
+\frac{e_0^4N_f}{48\pi^2}(|\bar z_1|^2+|\bar z_2|^2)\left\{2+3\ln\left[\frac{2e_0^2(|\bar z_1|^2+|\bar z_2|^2)}{\Lambda^2}\right]
\right\}+{\cal O}(N_f^2).
\end{equation}
The phase transition described by the above effective potential is certainly a first-order one.

To see how the second-order phase transition takes place, we perform a large $N_f$ expansion with the limit 
$N_f\to\infty$ being taken with $N_fe_0^2$ kept fixed. Up to order $1/N_f$ we have

\begin{eqnarray}
\label{Ueff-largeN}
&&U_{\rm eff}^{\rm ACL}(\bar z_1,\bar z_2)=(m^2-m^2_c)(|\bar z_1|^2+|\bar z_2|^2)+\left(\frac{u_0}{2}-\frac{8e_0^2}{N_f\pi^2}\right)
(|\bar z_1|^2+|\bar z_2|^2)^2\nonumber\\
&&-\frac{1}{12\pi}\left\{\left[m^2+u_0\left(3|\bar z_1|^2+|\bar z_2|^2\right)\right]^{3/2}
+\left[m^2+u_0\left(|\bar z_1|^2+3|\bar z_2|^2\right)\right]^{3/2}
+2(N_b-1)\left[m^2+u_0\left(|\bar z_1|^2+|\bar z_2|^2\right)\right]^{3/2}\right\}\nonumber\\
&&+{\cal O}\left(\frac{1}{N_f^2}\right),
\end{eqnarray}
\end{widetext}
where

\begin{equation}
m_c^2=\frac{e_0^4N_f}{24\pi^2}\left[3\ln\left(\frac{8\Lambda}{N_fe_0^2}\right)-1\right].
\end{equation}
The $(|\bar z_1|^2+|\bar z_2|^2)^2$-term is reminiscent of the Landau expansion of the dual formulation of 
the GL model of a superconductor,\cite{Kleinert-tric,Kleinert-tric-1} where a tricritical point was shown to exist. 
By defining a ``Ginzburg parameter'' $\kappa^2=u_0/(2e_0^2)$, we see that the effective potential 
(\ref{Ueff-largeN}) is stable only for $\kappa^2>8/(N_f\pi^2)$, in which case a second-order phase transition 
occurs.

The precise characterization of the change of behavior as a function of $N_f$ will be studied in the next Section 
by means of the RG.    

\section{Quantum critical behavior of the SU(2) algebraic charge liquid}

The $\beta$ functions at one-loop order in $d=4-\epsilon$ dimensions are 
easily calculated using standard methods.\cite{KSF,ZJ} The dimensionless couplings are given as 
a function of the momentum scale $\mu$ by 
$f=Z_A\mu^{-\epsilon}e_0^2$ and $g=Z_z^2\mu^{-\epsilon}u_0/Z_u$, where $Z_A$ and $Z_z$ are the wave function 
renormalizations for the gauge field and spinon field, respectively. $Z_u$ is obtained from the vertex function 
associated to spinon self-interaction. The renormalization factor $Z_A$ is obtained directly from the one-loop 
vacuum polarization. The $\beta$ functions $\beta_f\equiv\mu\partial f/\partial\mu$ and
$\beta_g\equiv\mu\partial g/\partial\mu$ are given by

\begin{equation}
\label{betaf}
\beta_f=-\epsilon f+\frac{4N_f+N_b}{3}f^2,
\end{equation}
and

\begin{equation}
\label{betag}
\beta_g=-\epsilon g-6fg+(N_b+4)g^2+6f^2,
\end{equation}
where $f$ and $g$ were rescaled as $f\to 8\pi^2f$ and $g\to 8\pi^2g$ in order to remove unnecessary geometrical factors. 
Note that at one-loop the number of fermion components $N_f$ appears only in Eq. (\ref{betaf}). Of course, for $N_f\to 0$ 
the above result coincides with the one for the Ginzburg-Landau superconductor.
\cite{Halperin-Lubensky-Ma,Folk,Kolnberger}         

For $N_b=2$, and provided $N_f\geq 4$, there are two fixed points $(g_\pm,f_*)$ at nonzero gauge coupling whose 
coordinates are given by

\begin{equation}
f_*=\frac{3\epsilon}{2(2N_f+1)},
\end{equation}
and

\begin{equation}
g_\pm=\frac{5+N_f\pm\sqrt{N_f^2+10N_f-56}}{6(2N_f+1)}\epsilon.
\end{equation}
From the above fixed points, only $(g_+,f_*)$ is infrared stable, 
thus corresponding to the quantum critical point where 
the critical exponents will be calculated. A schematic flow diagram is shown in Fig. 1. 
We see that the effect of Dirac fermions on the phase structure is 
quite remarkable, since neither $N_b$ nor $N_f$ need to be large in 
order to attain criticality. This behavior is in strong contrast with the one obtained 
in the limit $N_f\to 0$, where quantum criticality occurs only for $N_b>182.9$.\cite{Nogueira-Kragset-Sudbo}

\begin{figure}
\includegraphics[width=8cm]{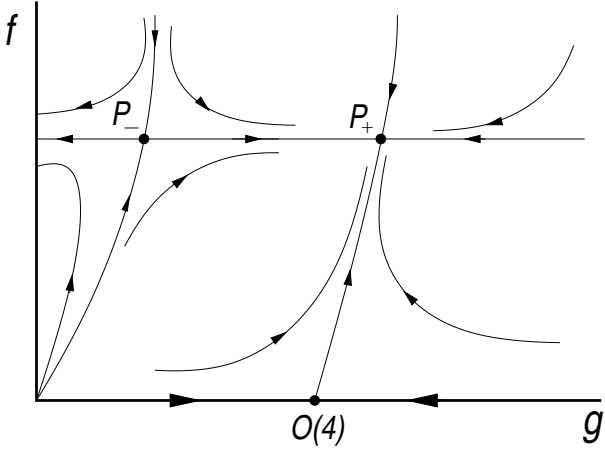}
\caption{Schematic flow diagram associated with the $\beta$ functions (\ref{betaf}) and (\ref{betag}) for 
$N_b=2$ and $N_f\geq 4$. In the figure $P_\pm=(g_\pm,f_*)$. Note that only $P_+$ is infrared stable, 
thus corresponding to a quantum critical point. For a vanishing gauge coupling we have a second-order 
phase transition governed by a $O(4)$ Wilson-Fisher fixed point. This fixed point is unstable 
for $f\neq 0$.} 
\end{figure} 

The case $N_f=4$ exhibits a peculiar critical behavior, as the calculations of the critical exponents will show. In fact, 
for $N_f=4$ the fixed points $(g_-,f_*)$ and $(g_+,f_*)$ coincide. 

The critical exponent $\nu$ is obtained through the calculation of the insertion of the 
operator $\sum_\alpha|z_\alpha|^2$ in the correlation function 
$\langle z_\alpha(x)z_\alpha^*(y)\rangle$.\cite{KSF,ZJ} This leads to new singularities, so that 
we have to introduce a new renormalization constant, $Z_2$, associated with this operator 
insertion. In pure scalar theories $Z_2$ is very simply calculated by 
differentiating the two-point function at zero momentum with respect to the mass. However, 
in general is better to perform the renormalization at nonzero momenta in order to avoid 
certain infrared divergences, especially when the scalar fields are coupled to a gauge field, like 
the case studied here. The standard way to calculate $\nu$ is to compute the following RG 
function at the infrared stable fixed point:

\begin{equation}
\label{gamma2}
\gamma_2\equiv\mu\frac{\partial\ln(Z_2/Z_z)}{\partial\mu},
\end{equation}
which leads to

\begin{equation}
\frac{1}{\nu}=2+\eta_2,
\end{equation}
where $\eta_2$ is the value of $\gamma_2$ at the infrared stable fixed point. 
At one-loop order, we have

\begin{equation}
\gamma_2=3f-(N_b+1)g.
\end{equation}
By inserting $f_*$ and $g_+$ for $N_b=2$ in the above equations and expanding up to first order 
in $\epsilon$, we obtain

\begin{equation}
\label{nu}
\nu=\frac{1}{2}+\frac{N_f-4+\sqrt{N_f^2+10N_f-56}}{4(1+2N_f)}~\epsilon+
{\cal O}(\epsilon^2).
\end{equation}

The critical exponent $\eta_N$ is also derived through a quadratic operator insertion. However, 
in this case there is a mass anisotropy involved. For a general $N_b$ the correlation 
function ${\cal G}(x)\equiv\langle{\bf n}(x)\cdot{\bf n}(0)\rangle$ is given in terms of 
spinon fields as

\begin{equation}
\label{G}
{\cal G}(x)=2\langle{\bf z}^*(x)\cdot{\bf z}(0)~{\bf z}(x)\cdot{\bf z}^*(0)\rangle
-\frac{2}{N_b}\langle|{\bf z}(x)|^2|{\bf z}(0)|^2\rangle.
\end{equation}
For $N_b=2$ we have simply

\begin{eqnarray}
{\cal G}(x)&=&2\langle z_1^*(x)z_1(0)z_2(x)z_2^*(0)\rangle+
2\langle z_1(x)z_1^*(0)z_2^*(x)z_2(0)\rangle
\nonumber\\ 
&-&2\langle|z_1(x)|^2|z_2(0)|^2\rangle+\langle|z_1(x)|^2|z_1(0)|^2\rangle\nonumber\\
&+&\langle|z_2(x)|^2|z_2(0)|^2\rangle.
\end{eqnarray}
The critical exponent $\eta_N$ is defined by the behavior of ${\cal G}(x)$ at the quantum critical point:

\begin{equation}
\label{G-scaling}
{\cal G}(x)\sim\frac{1}{|x|^{d-2+\eta_N}},
\end{equation}
where $\eta_N$ is given in terms of the crossover exponent $\varphi$ in Eq. (\ref{eta-N}).

We also note the following scaling behavior at the quantum critical 
point:

\begin{equation}
\label{density}
\left\langle\sum_\alpha|z_\alpha(x)|^2\sum_\beta|z_\beta(0)|^2\right
\rangle\sim\frac{1}{|x|^{d-2+\eta_4}},
\end{equation}
where 

\begin{equation}
\eta_4=d+2(1-1/\nu).
\end{equation}
Thus the scaling dimension of the operator $\sum_\alpha|z_\alpha(x)|^2$ is given by

\begin{equation}
{\rm dim}\left[\sum_\alpha|z_\alpha(x)|^2\right]=d-1/\nu.
\end{equation}
That $\eta_4$ depends only on $\nu$ is very easy to see, since it follows from the correlation function 
between operators $\sum_\alpha|z_\alpha|^2$, whose singular behavior 
leads to the renormalization constant $Z_2$ and consequently to $\nu$. 
From the expression (\ref{nu}) for the critical exponent $\nu$, we obtain

\begin{equation}
\label{eta4}
\eta_4=2-\frac{9+2\sqrt{N_f^2+10N_f-56}}{1+2N_f}~\epsilon+{\cal O}(\epsilon^2)
\end{equation}

The calculation of $\eta_N$ depends on the calculation of the crossover exponent $\varphi$. The calculation 
is facilitated by observing that the scaling dimension of each component of ${\bf n}$ is given by 
$d-\varphi/\nu$. In particular, we have 

\begin{equation}
{\rm dim}\left[|z_1|^2-|z_2|^2\right]=d-\varphi/\nu.
\end{equation}
Thus, $\varphi$ can be calculated from the insertions of $|z_1|^2$ and $|z_2|^2$ in the 
correlation function $G_{11}(x)\equiv\langle z_1(x)z_1^*(0)\rangle$.\cite{Amit}
This calculation is sketched in Appendix A.  By denoting those insertions by 
$G_{11,\alpha}$, where $\alpha=1,2$, we can obtain the crossover exponent from 
the renormalization of $G_{11,1}(x)-G_{11,2}(x)$.  
This procedure leads to a new renormalization constant $Z_2'$ and, analogously to Eq. (\ref{gamma2}), we 
can define the RG function

\begin{equation}
\label{gamma2-prime}
\gamma_2'\equiv\mu\frac{\partial\ln(Z_2'/Z_z)}{\partial\mu},
\end{equation}
which at the infrared stable fixed point yields

\begin{equation}
\label{eta2-prime}
\eta_2'=\frac{\varphi}{\nu}-2.
\end{equation}
By comparing the above equation with Eq. (\ref{eta-N}) we see that 
$\eta_N=d-2-2\eta_2'$. At one-loop order we have

\begin{equation}
\label{gamma2-prime-1}
\gamma_2'=3f-g,
\end{equation}
and therefore

%\begin{widetext}
\begin{equation}
\eta_N=2+\frac{\sqrt{N_f^2+10N_f-56}-5N_f-16}{3(1+2N_f)}~\epsilon+{\cal O}(\epsilon^2),
\end{equation}
%\end{widetext}
while the crossover exponent is given by

\begin{equation}
\varphi=1+\frac{5\left(N_f+\sqrt{N_f^2+10N_f-56}\right)-11}{12(1+2N_f)}~\epsilon
+{\cal O}(\epsilon^2).
\end{equation}
%\end{widetext}

\begin{table}
\caption{Values of the critical exponents for $N_b=2$ and different values of $N_f\geq 4$. We have 
set $\epsilon=1$.}
\begin{ruledtabular}
\begin{tabular}{ccccc}
$N_f$ & $\nu$ & $\eta_N$ & $\eta_4$ & $\varphi$\\
\hline
4 & $1/2$ & $1/3$ & 1 & $13/12$\\
5 & $0.62179$ & $0.61694$ & $0.38929$ & $1.27117$\\
6 & $0.66009$ & $0.75191$ & $0.33468$ & $1.3245$\\
7 & $0.68229$ & $0.84305$ & $0.3417$ & $1.35381$\\
8 & $0.69678$ & $0.90943$ & $0.36696$ & $1.37208$\\
9 & $0.70689$ & $0.96007$ & $0.39749$ & $1.38429$\\
10 & $5/7$ & $1$ & $3/7$ & $39/28$\\
11 & $0.71988$ & $1.0323$ & $0.45837$ & $1.39907$\\
20 & $0.73978$ & $1.17336$ & $0.64274$ & $1.41792$\\
50 & $0.74817$ & $1.27148$ & $0.83646$ & $1.42103$
\end{tabular}
\end{ruledtabular}
\end{table}

The critical exponents for three spacetime dimensions ($\epsilon=1$) are shown for several values of $N_f$ at Table I. As 
already mentioned, the case $N_f=4$ is peculiar. Indeed, for $N_f=4$ the critical exponents are rational and, moreover,  
it corresponds 
to the only case where $\eta_N<\eta_4$. Note that for $N_f=10$ the exponents are also rational, but in this case 
$\eta_N>\eta_4$, just like for all the other values of $N_f$. Interestingly, we have that $\eta_N>1$ for $N_f>10$. 
This result is consistent with the analysis at large $N_f$ and $N_b$ with $N_b/N_f$ arbitrary.\cite{Kaul-Sachdev} Indeed, 
in that case we find that $\eta_N>1$ if $N_f/N_b>3$ (if four-component Dirac fermions are used, the latter 
inequality becomes $N_f/N_b>3/2$). 

\section{Chiral symmetry breaking and chiral susceptibility}

We have already mentioned in the Introduction that CSB constitutes 
an important physical aspect of both algebraic Fermi and spin liquids. In this Section we want to investigate the 
effect of CSB on the ACL.\cite{Liu}  
This is typically a phenomenon that takes place at low energies. In the case 
of AFL and ASL it occurs for $|p|\ll N_fe_0^2$.\cite{Pisarski,Appelq} For the ACL this corresponds roughly 
to $|p|\ll\alpha\equiv(N_f+N_b/2)e_0^2$. In this regime the gauge field propagator is dominated by the one-loop vacuum 
polarization and is given by

\begin{equation}
D_{\mu\nu}(p)\approx\frac{16}{(2N_f+N_b)e_0^2|p|}\left(\delta_{\mu\nu}-\frac{p_\mu p_\nu}{p^2}\right),
\end{equation}
where we have used the Landau gauge and assumed a three-dimensional Euclidean spacetime. Following the 
standard literature,\cite{Pisarski,Appelq} 
we consider the one-loop fermion Green function with a dressed fermionic propagator and 
where the above gauge field propagator has been inserted. 
The inverse propagator is given by $G_\psi^{-1}(p)=i\slashchar{p}Z(p)+\Sigma(p)$ and the corresponding 
self-consistent equation reads

\begin{widetext}
\begin{equation}
G_\psi^{-1}(p)=i\slashchar{p}+\frac{16}{2N_f+N_b}\left\{
\int\frac{d^3k}{(2\pi)^3}\frac{\gamma_\mu
[\Sigma(k-p)+i(\slashchar{k}-\slashchar{p})Z(k-p)]\gamma_\mu}
{[Z^2(k-p)(k-p)^2+\Sigma^2(k-p)]|k|}
-\int\frac{d^3k}{(2\pi)^3}\frac{\slashchar{k}[\Sigma(k-p)+i(\slashchar{k}-
\slashchar{p})Z(k-p)]\slashchar{k}}{[Z^2(k-p)(k-p)^2+\Sigma^2(k-p)]|k|^3}\right\}.
\end{equation}
Note that the only difference with respect to the standard analysis is the coefficient in front of 
the curly brackets, which gets modified due to the spinon loop contribution to the vacuum polarization. We will 
set $N_b=2$, which is the case of physical interest to us. Futhermore, in order to simplify the analysis we will 
assume that $Z(p)\approx 1$.\cite{Appelq,Gusynin-1} After standard manipulations involving the algebra of the gamma matrices, 
we obtain the self-consistent equation for the self-energy:

\begin{equation} 
\Sigma(p)=\frac{32}{2N_f+1}\int\frac{d^3k}{(2\pi)^3}\frac{\Sigma(k)}
{[k^2+\Sigma^2(k)]|k+p|},
\end{equation} 
Integrating over the angles, we obtain

\begin{eqnarray}
\label{sc-int-sig}
\Sigma(p)&=&\frac{8}{(2N_f+1)\pi^2 |p|}\int_0^\alpha d|k|
\frac{|k|\Sigma(k)(|k|+|p|-|k-p|)}{k^2+\Sigma^2(k)}\nonumber\\
&=&\frac{16}{(2N_f+1)\pi^2 |p|}\left[\int_0^{|p|}d|k|\frac{k^2\Sigma(k)}{k^2+\Sigma^2(k)}
+|p|\int_{|p|}^\alpha d|k|\frac{|k|\Sigma(k)}{k^2+\Sigma^2(k)}\right],
\end{eqnarray}
\end{widetext}  
which can easily be converted to the differential equation

\begin{equation}
\label{d-Sig}
\frac{d}{dp}\left[p^2\frac{d\Sigma(p)}{dp}\right]=
-\frac{16}{\pi^2(2N_f+1)}\frac{p^2\Sigma(p)}{p^2+\Sigma^2(p)}.
\end{equation}
The above equation can be linearized in the regime $|p|\gg\Sigma(p)$ where it can be solved. We will omit 
the details here, since the analysis is well known.\cite{Appelq} The solution here is only slightly changed due to 
the spinon degrees of freedom. The result is a dynamically generated mass gap $\Sigma(0)\sim\langle\bar \psi\psi\rangle$ 
of the form

\begin{equation}
\Sigma(0)=\alpha\exp\left[-\frac{2\pi}{\sqrt{\frac{64}{\pi^2(2N_f+1)}-1}}\right].
\end{equation}
This mass generation is associated to the development of a chiral condensate, which signals the appearence of 
holon charge density waves in the system. The above mass gap vanishes for $N_f>N_{\rm ch}$, where

\begin{equation}
N_{\rm ch}=\frac{1}{2}\left(\frac{64}{\pi^2}-1\right)\approx 2.74,
\end{equation}
which is smaller than the corresponding value in absence of spinons.\cite{Note-CSB} 
Thus, CSB seems to occur only in the regime 
where no quantum critical point is found. Therefore, 
by accepting the calculations and underlying approximations employed so far in this paper, it seems that CSB 
does not spoil the quantum critical behavior. 

Another correlation function of interest is the chiral susceptibility

\begin{equation}
G_\chi(x)=\langle\bar \psi(x)\psi(x)\bar \psi(0)\psi(0)\rangle,
\end{equation}
whose scaling behavior for a large number of fermion components and $2+1$ dimensions has been 
calculated in Refs. \onlinecite{Gusynin-2} and \onlinecite{Franz}. At the quantum critical point the chiral susceptibility 
behaves like

\begin{equation}
\label{G-chiral}
G_\chi(x)\sim\frac{1}{|x|^{d-2+\tilde \eta_4}},
\end{equation}
which defines the critical exponent $\tilde \eta_4$. 
If fluctuations are ignored, we can write simply

\begin{equation}
G_\chi(x)\sim\frac{1}{|x|^{2(d-1)}},
\end{equation}
which leads to $\tilde \eta_4=d$.

We can obtain the anomalous dimension $\tilde \eta_4$ in $2+1$ dimensions 
for large $N_f$ and $N_b$, with $N_f/N_b$ arbitrary, directly from 
the result of Ref. \onlinecite{Franz} by the simple replacement $8/N\to 8/(N_f+N_b/2)$. The result agrees of 
course with the one obtained in Ref. \onlinecite{Kaul-Sachdev}, up to a 
trivial difference related to the fact that the latter reference works 
with two-component Dirac fermions.

Let us compute now $\tilde \eta_4$ for the $SU(2)$ ACL in first-order in $\epsilon$. 
Similarly to the case of four-spinon correlation functions, the calculation can be done by computing the insertion 
of the operator $\bar \psi\psi$ in the correlation function $G_\psi(x)$. This amounts in computing the 
scaling dimension of the operator $\bar \psi\psi$, i.e., 

\begin{equation}
\label{dim-chiral}
{\rm dim}[\bar \psi\psi]=d-1-\tilde \eta_2,
\end{equation}
where $\tilde \eta_2$ is the fixed point value of the RG function related to mass renormalization. Thus, 
the easiest way to compute the above scaling dimension is by adding a bare mass term in the Lagrangian 
${\cal L}_f$. The calculation of the needed renormalization constant is standard and can be found 
in quantum field theory textbooks.\cite{ZJ} We obtain at one-loop the result $\tilde \eta_2=3f_*$. From Eqs. 
(\ref{G-chiral}) and (\ref{dim-chiral}), we obtain

\begin{equation}
\tilde \eta_4=d-2\tilde \eta_2,
\end{equation}
and therefore,

\begin{equation}
\tilde \eta_4=2\left[2-\frac{N_f+5}{2N_f+1}\epsilon+{\cal O}(\epsilon^2)\right].
\end{equation}
In particular, for $N_f=4$ and $\epsilon=1$ we obtain $\tilde \eta_4=2$. In this case we have 
that the chiral susceptibility behaves at the quantum critical point as

\begin{equation}
G_\chi(x)\sim\frac{1}{|x|^{3}},
\end{equation}
which leads to a logarithmic behavior in momentum space.

\section{Conclusion and discussion} 

In this paper we have studied the quantum critical behavior of a field theory associated to the 
algebraic charge liquid 
(abreviated as ACL throughout this paper), which is a new type of 
quantum liquid proposed recently in Ref. \onlinecite{Kaul}. Our study was made in the framework 
of the $\epsilon$-expansion. To lowest order it leads to a quantum critical point provided 
the number of fermion components $N_f\geq 4$. Since we are considering a representation using 
four-component spinors, the physically relevant number of fermions is $N_f=2$.\cite{Note-3} 
Thus, in the framework of the present study, we are unable to find a deconfined quantum critical 
point for $N_f=2$. However, it is likely that an 
improved approximation will show that the theory is also critical for $N_f=2$. 
The point is that the $\epsilon$-expansion for the Lagrangian (\ref{L-boson}) should, in a nonperturbative 
framework, exhibit an infrared stable fixed point for all values of $N_b\geq 1$. The $\epsilon$-expansion 
cannot capture such a strong-coupling regime. At the moment there is no reliable nonperturbative 
scheme for this problem. As discussed in the introduction, a resummed three-loop 
$\epsilon$-expansion would provide the desirable controlled approximation. 
The inclusion of Dirac fermions make the $\epsilon$-expansion more reliable, since they essentially 
make the gauge coupling weaker. It is remarkable that in such a case it is not any longer necessary 
to have a large number of spinons to obtain a quantum critical point. Interestingly, the number of 
Dirac fermions does not need to be large either. Thus, we were able to explicitly exhibit a 
deconfined quantum critical point for an $SU(2)$ theory of bosonic spinons.

In order to convincingly show how the Dirac fermions make the full theory more well behaved, let us 
have a look at the two-loop $\beta$ function for the gauge coupling. This result can be easily 
derived by using the earlier two-loop result for the Lagrangian (\ref{L-boson}),\cite{Kolnberger} combined 
with the higher order result for the Lagrangian (\ref{L-fermion}).\cite{Gorishny,Gracey} The result is

\begin{equation}
\beta_f=-\epsilon f+\frac{N_b+4N_f}{3}f^2+2(N_b+2N_f)f^3,
\end{equation}
which has a nontrivial fixed point at

\begin{equation}
f_*=\frac{3\epsilon}{N_b+4N_f}\left[1-\frac{18}{N_b+4N_f}\epsilon+{\cal O}(\epsilon^2)\right].
\end{equation}
For $N_f=0$ we obtain that $f_*>0$ only for $N_b>18\epsilon$. Furthermore, in this 
limit the two-loop result for $\beta_g$ 
shows that we still have an infrared stable fixed point only for $N_b>182.9$. Thus, besides the two-loop 
result for $N_f=0$ being unable to produce a quantum critical point for low values of $N_b$, it introduces 
a new difficulty, since a nontrivial fixed point for $f$ does not any longer exists for all $N_b\geq 1$. 
However, the situation improves considerably if $N_f\neq 0$. Indeed, in this case we have that 
$f_*>0$ for $N_b>18\epsilon-4N_f$. Thus, for $N_b=2$ we should have $N_f>(9\epsilon-1)/2$, which after 
setting $\epsilon=1$ becomes $N_f>4$. While this clearly shows that the introduction of Dirac fermions 
improves the behavior of the $SU(2)$ antiferromagnet, it is a little bit frustrating to see that at 
two-loop no quantum critical point for $N_f=4$ is found anymore. However, we should keep in mind that 
the theory in absence of fermions is not being considered in a strong-coupling regime. Once this is 
done in some way, a critical regime even for values as low as $N_f=2$ should be expected to occur. Note that 
even if such a result is achieved, we may have to face another problem, namely, CSB. The calculation 
in Sect. IV indicates that CSB leads to gapped fermions for $N_f=2$. If this number is really 
correct, then there will be no deconfined quantum critical point anyway. Physically this means that 
CDW will screen the interspinon interaction in such a way as 
to make the spinon correlation length finite, and the only possible phase transition, if any, would 
be a first-order one. Adopting a more optimistic point of view, let us mention that there is strong 
evidence that the values of $N_{\rm ch}$ calculated using methods like the one we have used 
in Sect. IV tend to be overestimated.\cite{Appel,Hands,Note-4} Since the spinons contribute further to 
reduce the value of $N_{\rm ch}$, we may hope that the quantum critical point for $N_f=2$ will 
survive. 

Larger values of $N_f$ are physically relevant for multilayer systems. As discussed  
in the Introduction, this is precisely the case with the AFL.\cite{FT} However, we must be aware 
of the fact that in the case of the ACL additional gauge fields may play a role, so that 
the analysis made here would have to be modified accordingly. 

The main achievement of this paper was the computation of the anomalous dimensions of composite operators associated 
to the quantum critical behavior of the Lagrangian (\ref{L}). Particularly important was the calculation of the 
anomalous dimension $\eta_N$ of the N\'eel field. The results presented here constitute an explicit and well controlled 
example of deconfined quantum critical point for a system having $SU(2)$ invariance. The computed values of 
$\eta_N$ are substantially larger than the ones that would follow from a LGW analysis, reflecting a typical 
feature of the DQC scenario.    

\acknowledgments

The author would like to thank Hagen Kleinert and Zlatko Tesanovic for discussions. This work received 
partial support from the Deutsche Forschungsgemeinschaft (DFG), Grant No. KL 256/46-1.

\appendix \section{Calculation of $\eta_N$}

In this appendix we sketch the calculation of $\eta_N$ based on the insertion of the operators 
$|z_1|^2$ and $|z_2|^2$ in the correlation function $G_{11}(x)=\langle z_1(x)z_1^*(0)\rangle$. The calculation 
follows the method of Ref. \onlinecite{Amit}, i.e., it is based on dimensional regularization in the minimal 
subtraction scheme. 

\begin{figure}
\includegraphics[width=8cm]{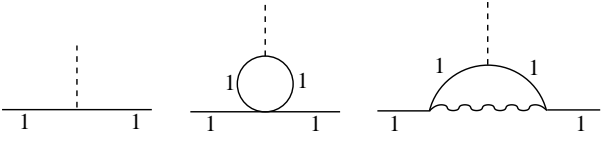}
\caption{One-loop Feynman diagrams representing the insertion of the operator $|z_1|^2$ (dashed line) in the correlation 
function $\langle z_1(x)z_1^*(0)\rangle$. Here the lines labeled $1$ or $2$ are associated to the 
fields $z_1$ and $z_2$ respectively. The wavy line represents the gauge field propagator.} 
\end{figure} 

In Fig. 2 the one-loop Feynman diagrams associated to the insertion of $|z_1|^2$ are shown. For the insertion 
of $|z_2|^2$, on the other hand, only one diagram is produced at one-loop (Fig. 3). 

The renormalization constant $Z_2'$ is determined by the normalization condition in momentum space

\begin{equation}
\label{R-cond-Z2p}
Z_2'\left[G_{11,1}^{-1}(p_1,p_2;p_3)-G_{11,2}^{-1}(p_1,p_2;p_3)\right]|_{\rm SP}=1,
\end{equation}
where $p_3=-(p_1+p_2)$ and SP denotes the symmetry point defined by

\begin{equation}
p_i\cdot p_j=\frac{\mu^2}{4}(4\delta_{ij}-1).
\end{equation} 

We will also need to compute $Z_z$, which can be determined from the normalization condition

\begin{equation}
\label{R-cond-Z}
\left.Z_z\frac{\partial G_{11}^{-1}(p)}{\partial p^2}\right|_{p^2=\mu^2}=1.
\end{equation}

A straightforward way to understand the insertion of the operators $|z_1|^2$ and $|z_2|^2$ is 
by adding a source term of the form 

\begin{equation}
\label{source}
{\cal L}_{\rm source}=r_1(x)|z_1(x)|^2+r_2(x)|z_2(x)|^2
\end{equation}
to the Lagrangian. Thus, 

\begin{equation}
G_{11,\alpha}(x;x')=\frac{\delta}{\delta r_\alpha(x')}G_{11}(x).
\end{equation}

\begin{figure}
\includegraphics[width=3cm]{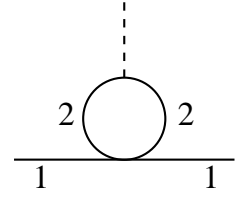}
\caption{For the insertion of the operator $|z_2|^2$ (dashed line) in the correlation 
function $\langle z_1(x)z_1^*(0)\rangle$ only one diagram contributes.} 
\end{figure}

Note that both $Z_z$ and $Z_2'$ are gauge dependent. However, this gauge dependence can 
be calculated exactly from the Ward identities.\cite{ZJ} It can be shown that $Z_z$ and $Z_2'$ have 
both the same gauge dependence to all orders in perturbation theory. Thus, the ratio 
$Z_2'/Z_z$ is gauge independent, which should be expected physically, since it leads to the anomalous dimension 
of the gauge invariant operator $|z_1|^2-|z_2|^2$. A similar argument holds for $Z_2/Z_z$, which leads 
to the critical exponent $\nu$. Since $Z_2'/Z_z$ is gauge independent, any gauge can be fixed to calculate 
$Z_z$ and $Z_2'$. Here we are going to use the Feynman gauge, so that the gauge field propagator is given 
in imaginary time by

\begin{equation}
D_{\mu\nu}(p)=\frac{\delta_{\mu\nu}}{p^2}.
\end{equation}

At the quantum critical point and at one-loop order, we have

\begin{eqnarray}
G_{11}^{-1}(p)&=&p^2-e_0^2\int\frac{d^dk}{(2\pi)^d}\frac{(2p_\mu-k_\mu)(2p_\nu-k_\nu)}{(p-k)^2}D_{\mu\nu}(k)
\nonumber\\
&=&p^2\left(1-\frac{e_0^2|p|^{-\epsilon}}{4\pi^2\epsilon}\right)+{\rm finite},
\end{eqnarray}
where we have made use of the integral

\begin{equation}
\label{int}
\int\frac{d^dk}{(2\pi)^d}\frac{1}{(p-k)^2k^2}=\frac{\Gamma(2-d/2)
\Gamma^2(d/2-1)}{(4\pi)^{d/2}\Gamma(d-2)}|p|^{d-4}
\end{equation}
and the usual properties of dimensional regularization.\cite{KSF,ZJ}
Next we use Eq. (\ref{R-cond-Z}) to obtain

\begin{equation}
Z_z=1+\frac{e_0^2\mu^{-\epsilon}}{4\pi^2\epsilon},
\end{equation}
which can be rewritten in terms of the renormalized dimensionless coupling up to one-loop accuracy as

\begin{equation}
Z_z=1+\frac{2f}{\epsilon}.
\end{equation}

The calculation of $Z_2'$ is made similarly. In this case standard manipulations in Eq. (\ref{R-cond-Z2p}) 
leads once more to integrals of the form (\ref{int}). An integral of the form 

\begin{equation}
\int\frac{d^dk}{(2\pi)^d}\frac{1}{(p_1-k)^2(p_2+k)^2k^2},
\end{equation} 
also appears in Eq. (\ref{R-cond-Z2p}). However, there is no need to worry with the above integral, since 
it does not have poles at $\epsilon=0$.  At the end we obtain the result

\begin{equation}
\label{Z2-p-result}
Z_2'=1+\frac{g}{\epsilon}-\frac{f}{\epsilon}.
\end{equation}
Therefore, 

\begin{equation}
\frac{Z_2'}{Z_z}=1+\frac{g}{\epsilon}-\frac{3f}{\epsilon},
\end{equation}
which leads to Eq. (\ref{gamma2-prime-1}).

The renormalization constant $Z_2'$ can also be obtained by 
computing $G_{11,\alpha}$ assuming that the sources in Eq. (\ref{source}) are 
$x$-independent. In this case, by setting $r_0=0$ we have

\begin{eqnarray}
&&G_{11}^{-1}(p)=p^2+r_1+(2u_0+e_0^2)\int\frac{d^dk}{(2\pi)^d}\frac{1}{k^2+r_1}
\nonumber\\
&&+u_0\int\frac{d^dk}{(2\pi)^d}\frac{1}{k^2+r_2}+2e_0^2\int\frac{d^dk}{(2\pi)^d}\frac{r_1-p^2}{[(p-k)^2+r_1]k^2}
\nonumber\\
&&-2e_0^2\int\frac{d^dk}{(2\pi)^d}\frac{1}{k^2}.
\end{eqnarray}
Thus, 

\begin{eqnarray}
&&G_{11,1}^{-1}(0,p;-p)=\frac{\partial}{\partial r_1}G_{11}^{-1}(p)\nonumber\\
&&=1-(2u_0+e_0^2)\int\frac{d^dk}{(2\pi)^d}\frac{1}{(k^2+r_1)^2}\nonumber\\
&&+2e_0^2\int\frac{d^dk}{(2\pi)^d}\frac{1}{[(p-k)^2+r_1]k^2}\nonumber\\
&&-2e_0^2p^2\int\frac{d^dk}{(2\pi)^d}\frac{1}{[(p-k)^2+r_1]^2k^2},
\end{eqnarray}
and
\begin{equation}
G_{11,2}^{-1}(0,p;-p)=\frac{\partial}{\partial r_2}G_{11}^{-1}(p)
=-u_0\int\frac{d^dk}{(2\pi)^d}\frac{1}{(k^2+r_2)^2}.
\end{equation}
The renormalization constant $Z_2'$ follows from

\begin{equation}
\lim_{p\to 0}\left[G_{11,1}^{-1}(0,p;-p)-G_{11,2}^{-1}(0,p;-p)\right]\left.\right|_{r_1=r_2=\mu^2}
=Z_2'^{-1},
\end{equation}
and the result (\ref{Z2-p-result}) is obtained once more.

In order to obtain $\eta_N$, we first note that the renormalized direction field, which is a composite operator, 
is given by

\begin{equation}
{\bf n}_R=\frac{Z_2'}{Z_z}{\bf n}.
\end{equation}
Since ${\bf n}_R$ is finite, its dimension is just given by dimenional analysis, i.e.,  
${\rm dim}[{\bf n}_R]=d-2$. Near the quantum critical point, we have that 
$Z_2'/Z_z\sim\mu^{\eta_2'}$, where $\eta_2'$ is the value of $\gamma_2'$ [recall Eq. (\ref{gamma2-prime})] 
at the infrared stable fixed point. Therefore, the scaling dimension of the N\'eel field is given by 

\begin{equation}
{\rm dim}[{\bf n}]=d-2-\eta_2'.
\end{equation}
The anomalous dimension $\eta_N$ is defined by the scaling behavior (\ref{G-scaling}). Thus, 
we also have that

\begin{equation}
{\rm dim}[{\bf n}]=\frac{d-2+\eta_N}{2},
\end{equation}
and therefore we obtain

\begin{equation}
\eta_N=d-2-2\eta_2'.
\end{equation}
Using Eq. (\ref{eta2-prime}) we obtain the form given in Eq. (\ref{eta-N}).


\begin{thebibliography}{100}
\bibitem{Note-2} Note a difference in notation with respect to Ref. \onlinecite{Kleinert-Nogueira-2}. There 
$N/2$ complex bosonic fields are considered, implying that the value $N=2$ actually corresponds to 
$N_b=1$ in our case.
\bibitem{Note-3} This corresponds to $N_f=2$ in Ref. \onlinecite{Kaul}, where two-component Dirac spinors 
are used. We have chosen to use four-component spinors because such a representation seems to be very natural 
in the ASL and AFL cases.\cite{Rantner,FT,Herbut-AFL,Kim,Wen-RMP} Furthermore, it facilitates the 
expansion around $d=4$, since at this dimensionality the natural representation is the one featuring four-component 
Dirac spinors.
\end{thebibliography}
\end{document}